\documentstyle[aps,prb,epsf]{revtex}

\begin{document}
\twocolumn[\hsize\textwidth\columnwidth\hsize\csname@twocolumnfalse\endcsname
\title{Role of in-plane dissipation in dynamics of Josephson lattice 
in high-temperature superconductors}
\author{A. E. Koshelev}
\address{Materials Science Division, Argonne National Laboratory, \\
Argonne, Illinois 60439}
\date{\today}
\maketitle

\begin{abstract}
We calculate the flux-flow resistivity of the Josephson vortex lattice
in a layered superconductor taking into account both the inter-plane and
in-plane dissipation channels.  We consider the limiting cases of small fields
(isolated vortices) and high fields (overlapping vortices).  In the
case of the dominating in-plane dissipation, typical for
high-temperature superconductors, the field dependence of flux-flow
resistivity is characterized by {\it three} distinct regions.  As
usual, at low fields the flux-flow resistivity grows linearly with
field.  When the Josephson vortices start to overlap the flux-flow
resistivity crosses over to the regime of {\it quadratic} field
dependence.  Finally, at very high fields the flux-flow resistivity
saturates at the c-axis quasiparticle resistivity.  The intermediate
quadratic regime indicates dominant role of the in-plane dissipation
mechanism.  Shape of the field dependence of the flux-flow resistivity
can be used to extract both components of the quasiparticle
conductivity.
\end{abstract}
\tighten

\vskip.2pc]
\narrowtext

Stack of low dissipative Josephson junctions formed by the atomic layers
of high-temperature superconductors \cite{IntrJos} represents
nonlinear system with unique dynamic properties.  Magnetic field
applied along the layers creates the lattice of Josephson vortices
(JVs).\cite{JosLat}  Transport properties of this state are
determined by dynamics of the Josephson lattice.  Two distinct regimes
exist depending on strength of applied magnetic field.  At low fields
the Josephson vortices are isolated and form a triangular lattice,
strongly stretched along the direction of layers.  An isolated JV is
characterized by the nonlinear core, the region within which the phase
difference between the two central layers sweeps from $0$ to $2\pi$.  The
core size is given by the Josephson length, $\gamma s$, where
$\gamma$ is the anisotropy of the London penetration depth and $s$ is
the interlayer spacing.  This regime of a dilute lattice is
characterized by the linear field dependence of flux-flow resistivity
$\rho_{ff}\propto B$.  The linear flux-flow branch in Bi$_2$Sr$_2$CaCu$_2$O$_8$
(BSCCO) at small fields has been observed experimentally.\cite{LeeAPL95,HechPRB97}

When magnetic field exceeds the crossover field, $B_{cr}=\Phi_0/\pi
\gamma s^2$, the cores of JVs start to overlap and a dense Josephson
lattice is formed, in which JVs fill all layers.\cite{DenseJosLat} In
this regime the linear field dependence of the flux-flow resistivity
breaks down.  Further field behavior depends on the mechanism of
dissipation.

Moving Josephson vortices generate both in-plane and inter-plane
electric fields, which induce dissipative quasiparticle currents. 
Usually only dissipation due to the tunneling of quasiparticles
between the layers is taken into account in calculation of the
viscosity of JVs.\cite{ClemCoffPRB90} However in the high-T$_{c}$
superconductors the in-plane quasiparticle conductivity $\sigma _{ab}$
is strongly enhanced in superconducting state as compared to the normal
conductivity due to reduction of phase space for scattering
\cite{SigmaQ,SigmaQBSCCO}, while the c-axis component $\sigma _{c}$
rapidly decreases with temperature in superconducting
state.\cite{Sigmac} Below the transition temperature the anisotropy of
dissipation $\sigma _{ab}/\sigma _{c}$ becomes larger than the
superconducting anisotropy $\gamma ^{2}$.  This leads to dominating
role of the in-plane dissipation in dynamics of the Josephson lattice.

In this Letter we calculate the field dependence of the flux-flow
resistivity $\rho_{ff}(B)$ taking into account both the in-plane and
inter-plane dissipation channels.  We separately consider the regimes of
small and high fields.  The flux flow resistivity at high fields, taking 
into account the in-plane dissipation, has 
been studied before in Ref.\ \onlinecite{BulPRB96}.
For the case of purely c-axis dissipation linear growth of
the flux-flow resistivity saturates at the c-axis quasiparticle
resistivity $\rho_{c}$ when magnetic field exceeds the crossover
field $B_{cr}$.  Dominating in-plane dissipation leads to qualitative change of
the shape of $\rho_{ff}(B)$.  In this case $\rho_{ff}$ also increases
linearly at small fields.  The slope of this dependence is mainly
determined by $\sigma_{ab}$ and at $B\approx B_{cr}$ the resistivity $\rho_{ff}$ is
still much smaller than $\rho_c$.  At $B\approx B_{cr}$ the field
dependence of $\rho_{ff}$ crosses over to even faster, {\em
quadratic}, dependence.  Only at significantly higher field, $B\approx
\sqrt{\sigma_{ab}/(\gamma^2\sigma_{c})}B_{cr}$, $\rho_{ff}$ reaches
$\rho_c$.  Therefore the field dependence of $\rho_{ff}$ can be used to
extract both components of the quasiparticle conductivity.

Dynamics of the moving Josephson lattice is governed by the coupled
Sine-Gordon equations for the interlayer phase
differences.\cite{CoupSinGord} The equations taking into account
in-plane dissipation have been derived in Refs.\ \onlinecite{BulPRB96,ArtJETPL97}. 
Consider a layered superconductor in magnetic field applied along the
layers ($y$ direction) and carrying transport current along the
$c$-axis ($z$ direction).  We express fields and currents in terms of
the gauge invariant phase difference between the layers $\theta
_{n}=\phi _{n+1}-\phi _{n}-\frac{2\pi s}{\Phi _{0}}A_{z}$ and the
in-plane superconducting momentum $p_{n}=\nabla _{x}\phi
_{n}-\frac{2\pi }{\Phi _{0}}A_{x}$.  The local magnetic field $B_{n}$
between the layers $n$ and $n+1$ can be expressed as
\begin{equation}
B_{n}(x)=\frac{\Phi _{0}}{2\pi s}\left( \frac{\partial \theta _{n}}{\partial
x}-p_{n+1}+p_{n}\right) .  \label{Hz}
\end{equation}
The components of electric field can be approximately represented as
\begin{equation}
E_{x}\approx \frac{\Phi _{0}}{2\pi c}\frac{\partial p_{n}}{\partial t}
;\;E_{z}\approx \frac{\Phi _{0}}{2\pi cs}\frac{\partial \theta _{n}}{
\partial t}.  \label{ExEz}
\end{equation}
These expressions are valid assuming fast equilibration of the order
parameter inside the layers.  More general situations have been considered in
Refs.\ \onlinecite{BulPRB96,KoyPRB96,ArtPRL97,RynJL97}.  The components
of electric current, $j_{x}$ and $j_{z}$, consist of the quasiparticle
and superconducting contributions
\begin{eqnarray}
j_{x} &=&\sigma _{ab}\frac{\Phi _{0}}{2\pi c}\frac{\partial p_{n}}{\partial t
}+\frac{c\Phi _{0}}{8\pi ^{2}\lambda _{ab}^{2}}p_{n},  \label{jx} \\
j_{z} &=&\sigma _{c}\frac{\Phi _{0}}{2\pi cs}\frac{\partial \theta _{n}}{
\partial t}+j_{J}\sin \theta _{n}.  \label{jz}
\end{eqnarray}
where $\sigma _{ab}$ and $\sigma _{c}$ are the components of the
quasiparticle conductivity, $\lambda _{ab}$ and $ \lambda _{c}$ are the
components of the London penetration depth, and $j_{J}=c\Phi _{0}/(8\pi
^{2}s\lambda _{c}^{2})$ is the Josephson current density.  Note that the
in-plane current is actually concentrated inside the superconducting
layers and physically meaningful quantities are the two-dimensional
current densities $J_{xn}$ in the layers.  To be precise, the bulk current in Eq.\
(\ref{jx}) is defined at discrete points $z_{n}=ns$ as
$j_{x}(z_{n})\equiv J_{xn}/s$.  Using above relations we rewrite the $z$
and $x$ components of the Maxwell equation \[\frac{4\pi }{c}{\bf
j}+\frac{\partial {\bf D}}{\partial t}=\nabla \times {\bf B}\] as
\begin{eqnarray}
&&\frac{2\sigma _{c}\Phi _{0}}{c^{2}s}\frac{\partial \theta _{n}}{\partial t}+
\frac{4\pi }{c}j_{J}\sin \theta _{n}+\frac{\varepsilon _{c}\Phi _{0}}{2\pi
c^{2}s}\frac{\partial ^{2}\theta _{n}}{\partial t^{2}} =\frac{\partial
B_{n}}{\partial x},  \label{Maxwz} \\
&&\frac{2\sigma _{ab}\Phi _{0}}{c^{2}}\frac{\partial p_{n}}{\partial t}+\frac{
\Phi _{0}}{2\pi \lambda _{ab}^{2}}p_{n} =-\frac{B_{n}-B_{n-1}}{s}.
\label{Maxwx}
\end{eqnarray}
In the second equation we replaced $\partial B/\partial z$ by the
discrete derivative $(B_{n}-B_{n-1})/s$.  We also neglected the
in-plane displacement current $\partial D_{x}/dt$, because typical
frequencies involved in Josephson dynamics are much smaller than the
in-plane plasma frequency $ c/\lambda _{ab}$.  Eqs.  (\ref{Hz}),
(\ref{Maxwz}), and (\ref{Maxwx}) give closed system which describes
dynamics of the phases $\theta _{n}(x,t)$, fields $ B_{n}(x,t)$ and
momenta $p_{n}(x,t)$.  The moving lattice generates both in-plane and
c-axis electric fields (\ref{ExEz}) leading to dissipation. 
The rate of energy dissipation $W$ per unit volume is given by
\begin{equation}
W=\left( \frac{\Phi _{0}}{2\pi c}\right) ^{2}\left [\frac{
\sigma _{c}}{s^{2}}\left \langle \left( \frac{\partial \theta _{n}}{\partial t}\right)
^{2}\right \rangle+\sigma _{ab}\left \langle\left( \frac{\partial p_{n}}{\partial t}\right) 
^{2}\right \rangle\right ].
\label{Dissip}
\end{equation}
For the steady state motion with small velocity $v$ the phase differences
vary in space and time as $\theta _{n}(x,t)=\theta _{n}^{(0)}(x-vt)$,
where $\theta _{n}^{(0)}(x)$ is the static phase distribution, and
Eq.\ (\ref {Dissip}) can be rewritten as
\[
W=\eta _{J} v^{2},
\]
where
\begin{equation}
\eta _{J}=\left( \frac{\Phi _{0}}{2\pi c}\right) ^{2}\left[ 
\frac{\sigma _{c}}{s^{2}}\left\langle \left( \frac{\partial \theta _{n}^{(0)}}{\partial x}\right)
^{2}\right\rangle +\sigma _{ab}\left\langle \left( \frac{\partial p_{n}^{(0)}
}{\partial x}\right) ^{2}\right\rangle \right]   \label{Viscos}
\end{equation}
is the linear viscosity coefficient of the lattice per unit volume and $
\left\langle \ldots \right\rangle $ means averaging with respect to $x$ and $
n$. The flux-flow resistivity $\rho _{Jff}$ is connected with $\eta _{J}$ by
relation $\rho _{Jff}=B^{2}/(c^{2}\eta _{J})$.

Consider the regime of small fields, $B\ll B_{cr}=\Phi _{0}/(\pi \gamma
s^{2})$.  In this regime the JVs are isolated and dissipation is
concentrated in the vicinity of nonlinear vortex cores.  In this case
$\eta _{J}$ is proportional to the field $\eta _{J}=B\eta _{Jv}/\Phi
_{0}$, where $ \eta _{Jv}$ is the viscosity coefficient of an individual
JV per unit length.  The viscosity coefficient due to the c-axis
dissipation has been calculated by Clem and Coffey.\cite{ClemCoffPRB90}
Similar problems of viscous friction have been studied for an Abrikosov
vortex (see, e.g., review \onlinecite{AVvisc}) and for a Josephson
vortex in a single junction.\cite{JVvisc}

In the vicinity of the vortex cores one can neglect screening
effects and express the phase differences $\theta _{n}(x)$ and momenta
$p_{n}(x)$ via the in-plane phases $ \phi _{n}(x)$, $\theta
_{n}^{(0)}\approx \phi _{n+1}-\phi _{n}$, $p_{n}^{(0)}\approx \nabla _{x}\phi
_{n}$ (we are using the gauge ${\rm div}{\bf A}=0$).  Numerically accurate
phase distribution $\phi _{n}(x)$ in the vicinity of
the vortex core was obtained in Ref.\ \onlinecite{KinkWalls},
\begin{eqnarray}
\phi _{n}(u) &\approx &\arctan \frac{n-1/2}{u}+\frac{0.35(n-1/2)u}{\left(
(n-1/2)^{2}+u^{2}+0.38\right) ^{2}}   \nonumber\\
&+&\frac{8.81(n-1/2)u\left( u^{2}-(n-1/2)^{2}+2.77\right) }{\left(
(n-1/2)^{2}+u^{2}+2.02\right) ^{4}}.  \label{NumPhases}
\end{eqnarray}
with $u\equiv x/\gamma s$.\cite{Note-phin} We can represent now the
viscosity coefficient $\eta _{Jv}$ as
\[
\eta _{Jv}=\frac{1}{\gamma s^{2}}\left( \frac{\Phi _{0}}{2\pi c}\right)
^{2}\left[ C_{c }\sigma _{c}+C_{ab }\frac{\sigma _{ab}}{\gamma ^{2}}\right] ,
\]
with
\begin{eqnarray*}
C_{c } &=&\sum_{n=-\infty}^{\infty}\int_{-\infty}^{\infty} du\left( \frac{\partial \left( \phi _{n+1}-\phi
_{n}\right) }{\partial u}\right) ^{2}, \\
C_{ab } &=&\sum_{n=-\infty}^{\infty}\int_{-\infty}^{\infty} du\left( \frac{\partial ^{2}\phi _{n}}{\partial
u^{2}}\right) ^{2}.
\end{eqnarray*}
Using the phase distribution (\ref{NumPhases}) we compute $C_{c }\approx 9.0$
and $C_{ab }\approx 2.4$. Finally, we obtain the following result for the
flux-flow resistivity at small fields $\rho _{Jff}=\Phi _{0}B/(c^{2}\eta 
_{Jv})$
\begin{equation}
\rho _{Jff}\approx \frac{4.4\gamma s^{2}B}{\Phi _{0}\left( 
\sigma_{c}+0.27
\sigma_{ab}/\gamma ^{2}\right) },\text{ at }B<B_{cr}
\label{rhoffLow}
\end{equation}
The case of dominating c-axis dissipation ($\sigma_{ab}\ll
\gamma^{2}\sigma_{c}$) has been considered by Coffey and Clem
\cite{ClemCoffPRB90}.  They obtain the coefficient 2.8 instead of 4.4
using the approximate phase distribution.

\begin{figure}
\epsfxsize=3.25in \epsffile{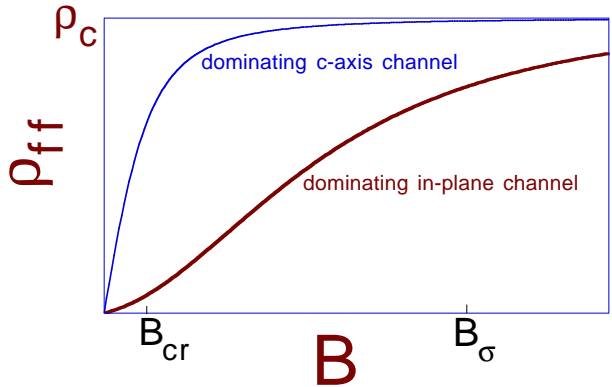} 
\caption{Schematic field
dependence of the flux-flow resistivity for the cases of dominating
c-axis dissipation channel ($\sigma_{ab}/\sigma_{c}\ll \gamma^2$) and
dominating in-plane dissipation channel ($\sigma_{ab}/\sigma_{c}\gg
\gamma^2$).  For strong in-plane dissipation the dependence
$\rho_{ff}(B)$ has pronounced upward curvature at $B\protect \gtrsim B_{cr}$
and approaches $\rho_{c}$ at the field $B_{\sigma}$ much larger than
the crossover field $B_{cr}$.}
\label{Fig-rhoff}
\end{figure}

Now we consider the regime of high fields, $B > B_{cr}$.  In this regime
we can obtain a simple analytical result using expansion with respect to
the Josephson current.\cite{BulPRB96} For the static lattice in the zero
order with respect to $j_{J}$ we have $B_{n}(x)=B$, $p_{n}(x)=0$, and
$\theta _{n}(x)=2 \pi sBx/\Phi _{0}+\pi n$.  The first iteration with
respect to $j_{J}\equiv \frac{ c\Phi _{0}}{8\pi ^{2}s\lambda _{c}^{2}}$
gives
\begin{eqnarray}
B_{n}(x) &=&B-\frac{\Phi _{0}^{2}}{16\pi ^{2}s^{2}\lambda _{c}^{2}B}\cos
\left( \frac{2\pi sB}{\Phi _{0}}x+\pi n\right)  \\
p_{n}(x) &=&\frac{\Phi _{0}}{\pi Bs^{3}\gamma ^{2}}\cos \left( \frac{2\pi sB
}{\Phi _{0}}x+\pi n\right)
\end{eqnarray}
Substituting expressions for $\theta _{n}(x)$ and $p_{n}(x)$ into Eq.\ (\ref
{Viscos}) we obtain
\begin{equation}
\eta _{J}=\left( \frac{\Phi _{0}}{\pi c\gamma s^{2}}\right) ^{2}\left[
\sigma _{c}\left( \frac{\pi s^{2}\gamma B}{\Phi _{0}}\right) ^{2}+\frac{
\sigma _{ab}}{2\gamma ^{2}}\right],
\label{etaJhigh}
\end{equation}
and
\begin{equation}
\rho _{Jff}=\frac{B^{2}}{ B^{2}+ B_{\sigma}^{2}}\rho 
_{c}, \; \; 
B_{\sigma}=\sqrt{\frac{\sigma_{ab}}{\sigma_{c}}}\frac{\Phi _{0}}{\sqrt{2}\pi \gamma
^{2} s^{2}},
\label{rhoffHigh}
\end{equation}
for $B>B_{cr}$.\cite{NoteComp} At $B\approx B_{cr}$ this result
approximately matches with Eq.\ (\ref{rhoffLow}).  Eqs.\
(\ref{rhoffLow}) and (\ref{rhoffHigh}) represent the main results of
this Letter.  We see that the shape of $\rho _{Jff}(B)$ is determined by
the ratio $\sigma_{ab}/(\gamma^2\sigma_{c})$.  In the high-temperature
superconductors typically $\sigma_{ab}/\sigma_{c}\approx \gamma ^{2}$
near the transition temperature.  However the ratio
$\sigma_{ab}/\sigma_{c}$ rapidly becomes much larger than $\gamma ^{2}$
with temperature decrease because of (i) significant enhancement of
$\sigma _{ab}$ due to the suppression of in-plane scattering of
quasiparticles \cite{SigmaQBSCCO} and (ii) fast decrease of $\sigma
_{c}$.\cite{Sigmac} This means that dependence $\rho _{c}\propto B^{2}$
holds in a wide field range $B_{cr}<B<B_{\sigma}$.  The field $B_{cr}$
is almost temperature independent and for optimally doped BSCC0 ($\gamma
\approx 500$) $B_{cr}\approx 0.5$T. The field $B_{\sigma}$ has strong
temperature dependence via $\sigma_{ab}(T)$ and $\sigma_{c}(T)$.  Taking
typical values for $T\approx 20$K, $\sigma_{c}=2\cdot
10^{-3}$(Ohm$\cdot$cm$)^{-1}$ and $\sigma_{ab}\approx
5\cdot10^{4}$(Ohm$\cdot$cm$)^{-1}$, we obtain estimate
$B_{\sigma}\approx 4$T. The field dependencies for the cases of
dominating c-axis dissipation and dominating in-plane dissipation are
sketched in Fig.\ \ref{Fig-rhoff}.  The shapes of the field dependencies
are qualitatively different for these two cases.  In particular, they
have opposite curvatures at $B\lesssim B_{cr}$.  Therefore, the shape of
$\rho_{Jff}(B)$ can be used to extract both components of the
quasiparticle conductivity.  At present, there are no published data for
the field dependence of the flux-flow resistivity in a wide field range. 
Dynamics of the Josephson lattice at high fields has been studied by G.\
Hechtfischer {\it et al.}\cite{HechPRL97} From the I-U curves presented
in this paper one can see that the slope $dI/dU$ for the first flux-flow
branch indeed has a strong field dependence with upward curvature in the
field range $2-3.5$ tesla.

In conclusion, we calculated the flux-flow resistivity of the
Josephson lattice at small and high fields and demonstrated that
strong in-plane dissipation qualitatively modifies its field
dependence.

I would like to thank V.\ Zavaritskii and R.\ Kleiner for helpful
discussions and D.\ Dom\'{i}nguez for attracting my attention to Ref.\
\onlinecite{BulPRB96}, in which dynamics of the Josephson lattice at
high fields has been studied.  This work was supported by the US DOE,
BES-Materials Sciences, under contract No.  W-31-109-ENG-38.  The author
also would like to acknowledge support from the Japan Science and
Technology Corporation, STA Fellowship 498051, and to thank National
Research Institute for Metals (NRIM) in Tsukuba for hospitality.

\begin{references}
\bibitem{Note-phin} Unfortunately, the formula (B12) in Ref.\ 
\protect\onlinecite{KinkWalls} for phase distribution $\phi_{n}$ was 
written incorrectly.  Eq.\ (\ref{NumPhases}) gives correct phase 
distribution, which was actually used in numerical calculations and is plotted 
in Fig.\ 8 of Ref.\ \protect\onlinecite{KinkWalls}.
\bibitem{NoteComp}Eqs.\ (\protect\ref{etaJhigh}) and
(\protect\ref{rhoffHigh}) are similar to Eq.\ (42) from Ref.\ \protect
\onlinecite{BulPRB96} for the flux flow conductivity except for the sign
in front of the in-plane term.
\end{references}
\end{document}